\documentclass{ws-p10x7}
\usepackage{amsmath}
\usepackage{amssymb}
\usepackage{euscript}
\usepackage{color}
\def\Color#1{\color[named]{#1}}
\def\Order#1{${\cal O}(#1)$}
\def\bbeta{\bar{\beta}}

\def\Ordpr#1{${\cal O}(#1)_{prag}$}

\def\bbeta{\bar{\beta}}
\def\hbeta{\hat{\beta}}

\newcommand{\Meu}{\EuScript{M}}

\newcommand{\sfac}{\mathfrak{s}}
\newcommand{\GeV}{\unskip\,\mathrm{GeV}}
\def\mathswitchr#1{\relax\ifmmode{\mathrm{#1}}\else$\mathrm{#1}$\fi}

\newcommand{\Pd}{\mathswitchr d}
\newcommand{\Pu}{\mathswitchr u}
\newcommand{\Ps}{\mathswitchr s}
\newcommand{\Pc}{\mathswitchr c}

\begin{document}

\title{Coherent Exclusive Exponentiation for Precision Monte Carlo 
Calculations of Fermion Pair Production $/$ Precision Predictions 
for (Un)stable $W^+W^-$ Pairs}

\author{B.F.L. Ward$^{a,b,c}$, S. Jadach$^{ a,b,d,e}$, W. P{\L}aczek$^{b,f}$, M. Skrzypek$^{b,e}$, \\ and Z. Was$^{b,e}$}

\address{$^A$Department of Physics and Astronomy, University of Tennessee,
Knoxville, TN 37996-1200, USA\\
$^B$TH Div., CERN, CH-1211 Geneva 23, Switzerland\\
$^C$SLAC, Stanford UNiversity, Stanford, CA 94309, USA\\
$^D$Theory Div., DESY, D-22603 Hamburg, Germany\\
$^E$Institute of Nuclear Physics, ul. Kawiory 26a, PL-30-055 Krakow, Poland\\
$^F$Institute of Computer Science, Jagellonian University, ul. Nawojki 11,
30-072 Krakow, Poland}


\author{UTHEP-00-0901, Sept., 2000}

\address{ Presented by B.F.L. Ward at ICHEP2000, Osaka, Japan, July 27, 2000}

\twocolumn[\maketitle\abstract{ 
We present the new Coherent Exclusive Exponentiation (CEEX),
in comparison to the older Exclusive Exponentiation (EEX)
and the semi-analytical Inclusive Exponentiation (IEX),
for the process $e^+e^-\rightarrow f\bar{f} +n\gamma$, $f=\mu,\tau,d,u,s,c,b$,
with validity for centre of mass energies 
from $\tau$ lepton threshold to 1~TeV.
We analyse $2f$ numerical results at the $Z$-peak, 189~GeV and 500~GeV.
We also present precision calculations of the signal processes 
$e^+e^-\rightarrow 4f$ in which the double resonant $W^+W^-$
intermediate state occurs using our YFSWW3-1.14 MC.
Sample $4f$ Monte Carlo data are explicitly illustrated in comparison to
the literature at LEP2 energies. 
These comparisons show that a TU
for the signal process cross section of $0.4\%$ is valid for the LEP2
200 GeV energy.  LC energy results are also shown.}]

\section{Introduction}
At the end of the LEP2 operation, the total cross section for 
the process $e^-e^+\to f\bar{f}+n\gamma$
will have to be calculated with the precision $0.2\%$ - $1\%$, depending on 
the event selection. In addition, the awarding of the 1999 Nobel
Prize to G. 't Hooft and M. Veltman
emphasises the importance of the on-going
precision studies of the Standard Model processes
$e^+e^- \to W^+W^- +n(\gamma)\to 4f+n(\gamma)$
at LEP2
energies, 
as well as the importance of the planned future
higher energy studies of such processes in
LC physics 
programs.
\par
In what follows, we present precision predictions
for both sets of processes, using our new 
coherent exponentiation (CEEX)~\cite{ceex} theory 
(${\cal KK}$ MC) for the former set and our older
and firmly established exclusive
exponentiation (EEX)~\cite{eex} theory (YFSWW3-1.14 MC~\cite{yfsww3})
for the latter set. Both CEEX and and EEX 
are based on the YFS exclusive exponentiation theory of
Yennie, Frautschi and Suura~\cite{yfs}. 
A detailed description~\cite{ceex,yfsww3,eex} of our two approaches to the
precision exponentiation theory may be found in Refs.~\cite{ceex,yfsww3,eex}.
As we indicate below, we have compared our ${\cal KK}$ MC calculations with
with EEX, its semianalytical 
partner IEX, and {\Color{Maroon}ZFITTER 6.21}~\cite{zfitter}
and we have compared our YFSWW3-1.14 MC
calculations with {\Color{Maroon}RacoonWW}~\cite{ditt} and with 
the Beenakker {\it et al.}\cite{been1} {\Color{Maroon}semi-analytical approach}.
\par
The paper proceeds as follows. In Sec. 2 we discuss the implementation of CEEX in our ${\cal KK}$ MC in relation to EEX.
In Sec. 3 we present some of its
new results for $2f+n(\gamma)$ processes at high energies.
In Sec. 4 we present the EEX theory
realization in our YFSWW3-1.14 MC. In Sec. 5 we present some of its
new results on $WW+n(\gamma)\rightarrow 4f+m(\gamma)$ processes
at high energies. Sec. 6 contains our summary remarks.\par

\section{${\cal KK}$ MC}

The main differences between CEEX and EEX are
best illustrated by focusing on the process of interest, which is
\begin{equation}
\label{eq1}
\begin{split}
& e^-(p_1,{\Color{Red}\lambda}_1)+e^+(p_2,{\Color{Red}\lambda}_2) \to 
 f(q_1,{\Color{Red}\lambda}'_1)+\bar{f}(q_2,{\Color{Red}\lambda}'_2)\\
&+{\Color{Maroon}\gamma}(k_1,{\Color{Black}\sigma}_1)+...+{\Color{Maroon}\gamma}(k_n,{\Color{Black}\sigma}_n).
\end{split}
\end{equation}
The respective
{\Color{Magenta} EEX} total cross section
{\Color{PineGreen}
\begin{equation}
\label{eq2}
\sigma = \sum\limits_{n=0}^\infty \;\;
          \int\limits_{m_{\Color{Maroon}\gamma}}
          d\Phi_{n+2}\; e^{Y({m_{\Color{Maroon}\gamma}})}
          D_n(q_1,q_2,k_1,...,k_n)
\end{equation}
}
corresponds to the attendant
{\Color{PineGreen}\Order{\alpha^1}} distributions $D_n$ as given 
in Ref.~\cite{eex} by formulas such as, for 
{\Color{Red}$n=0,1$}, $D_0 =  \bbeta_0$ and 
$D_1(k_1)     =  \bbeta_0 {\Color{Maroon}\tilde{S}}(k_1) +\bbeta_1(k_1)$, 
where the
{\Color{PineGreen}real soft factors} ${\Color{Maroon}\tilde{S}}(k)$
are defined as usual~\cite{eex}.
The important point is that the IR-finite building blocks $\bbeta_n$,
for example,
$\bbeta_0= \sum_{\Color{Red}\lambda} |\Meu^{\rm Born}_{\Color{Red}\lambda}|^2$,
in the multi-photon distributions are all
in terms of {\Color{PineGreen}$\sum\limits_{spin} |...|^2 $}{\Color{Red}!}
Here, {\Color{Black}${\Color{Red}\lambda} =$ fermion helicities} and {\Color{Black}${\Color{Black}\sigma} =$ photon helicity}.
{\Color{Blue}
{\Color{Magenta}In contrast}, in the analogous {\Color{Black}\Order{\alpha^1}} case of {\Color{Magenta}CEEX}
{\Color{PineGreen}
\begin{equation}
\begin{split}
\sigma &= \sum\limits_{n=0}^\infty\;
          \int\limits_{{ m_{\Color{Maroon}\gamma}}} d\Phi_{n+2}\\
       &  \sum\limits_{{\Color{Red}\lambda},{\Color{Black}\sigma}_1,...,{\Color{Black}\sigma}_n}
          |e^{B({m_{\Color{Maroon}\gamma}})}
          \Meu^{{\Color{Red}\lambda}}_{n,{\Color{Black}\sigma}_1,...,{\Color{Black}\sigma}_n}(k_1,...,k_n)|^2
\end{split}
\end{equation}
}
{\Color{Maroon}the differential distributions}
for {\Color{Maroon}$n=0,1$} photons are, for example,
{\Color{Magenta}$\Meu_{0}^{{\Color{Red}\lambda}} = \hbeta_0^{\Color{Red}\lambda},\quad {\Color{Red}\lambda}={\rm fermion~ helicities}$ and 
$\Meu^{\Color{Red}\lambda}_{1,{\Color{Black}\sigma}_1}(k_1) 
              = \hbeta^{\Color{Red}\lambda}_0 {\Color{Maroon}\sfac}_{{\Color{Black}\sigma}_1}(k_1)
               +\hbeta^{\Color{Red}\lambda}_{1,{\Color{Black}\sigma}_1}(k_1)$
,} with the {\Color{Red}IR-finite} building blocks {\Color{PineGreen}$\hbeta^{\Color{Red}\lambda}_0 = \big(e^{-B} \Meu^{\rm Born+Virt.}_{{\Color{Red}\lambda}}\big)\big|_{{\cal O}(\alpha^1)}$ and
$\hbeta^{{\Color{Red}\lambda}}_{1,{\Color{Black}\sigma}}(k)=\Meu^{\Color{Red}\lambda}_{1,{\Color{Black}\sigma}}(k) - \hbeta^{\Color{Red}\lambda}_0 {\Color{Maroon}\sfac}_{{\Color{Black}\sigma}}(k)$}.
Explicitly,{\Color{Orange} this time
everything is in terms of} {\Color{Magenta}$\Meu$-spin-amplitudes}!
{\Color{Black}This is the basic difference between EEX/YFS AND CEEX.}
{\Color{Maroon}Complete expressions for
spin amplitudes with CEEX exponentiation, $n_{\Color{Maroon}\gamma}$ arbitrary,
are given in  {\Color{PineGreen}Phys. Lett. {\bf B449},  97  (1999)} for the 
{\Color{PineGreen}\Order{\alpha^1}} case
and in {\Color{Magenta}CERN-TH/2000-087,UTHEP-99-09-01}, for the 
{\Color{Magenta}\Order{\alpha^2}} case, all are based on {\Color{Red}GPS}
spinor conventions as given in {\Color{Magenta}CERN-TH-98-235, hep-ph/9905452}.
}
}
\section{Results: CEEX}
\begin{figure}
\begin{center}
\setlength{\unitlength}{0.05mm}
\begin{picture}(850,1750)
\put(400,1750){\makebox(0,0)[b]{{\Color{Magenta} Physical Precision Tests}}}
\put( 400,1680){\makebox(0,0)[b]{ (a)}}
\put(400,800){\makebox(0,0)[b]{ (b)}}
\put(  -20,880){\makebox(0,0)[lb]{\epsfig{file=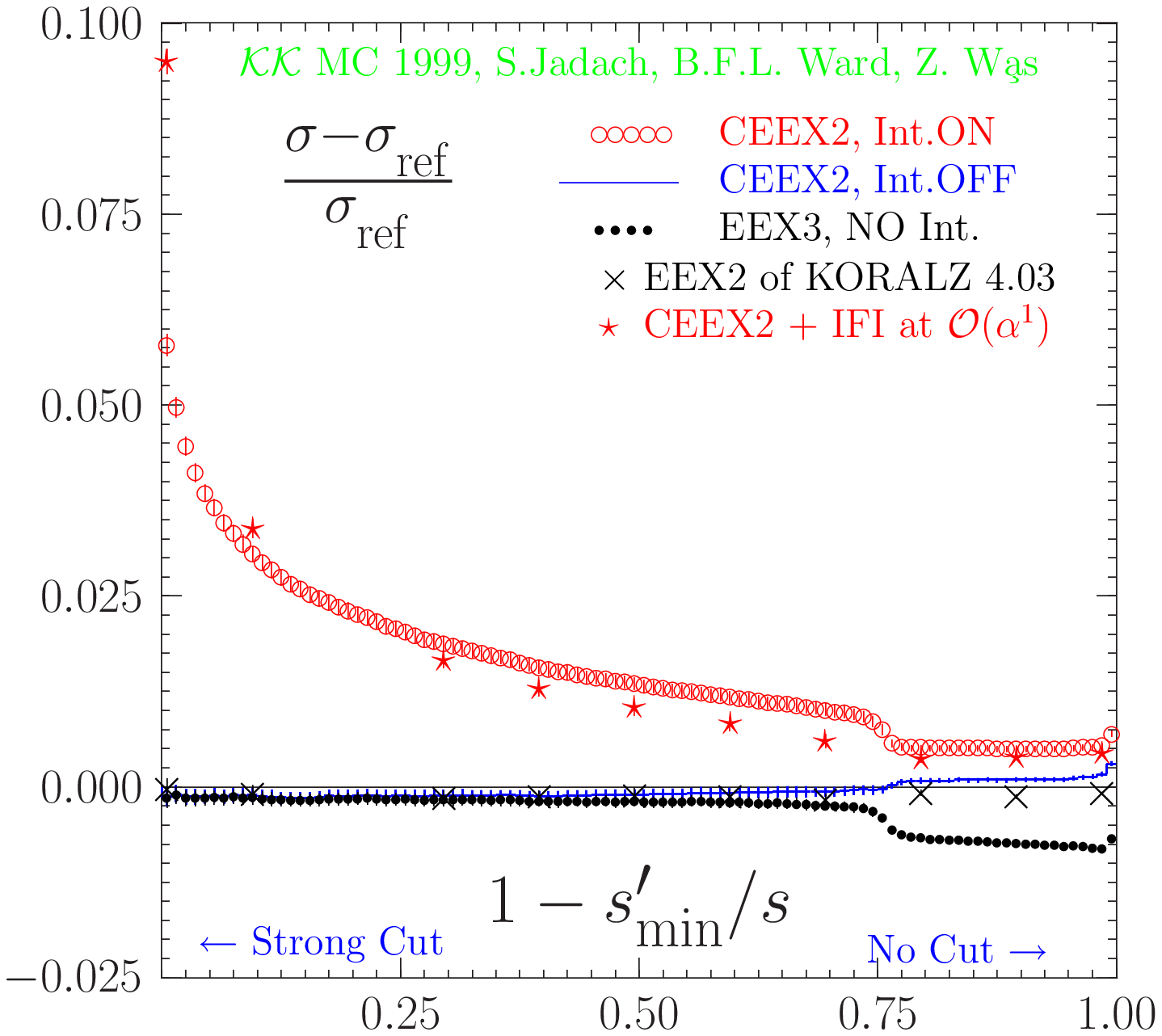,width=40mm,height=40mm}}}
\put(  -20, 0){\makebox(0,0)[lb]{\epsfig{file=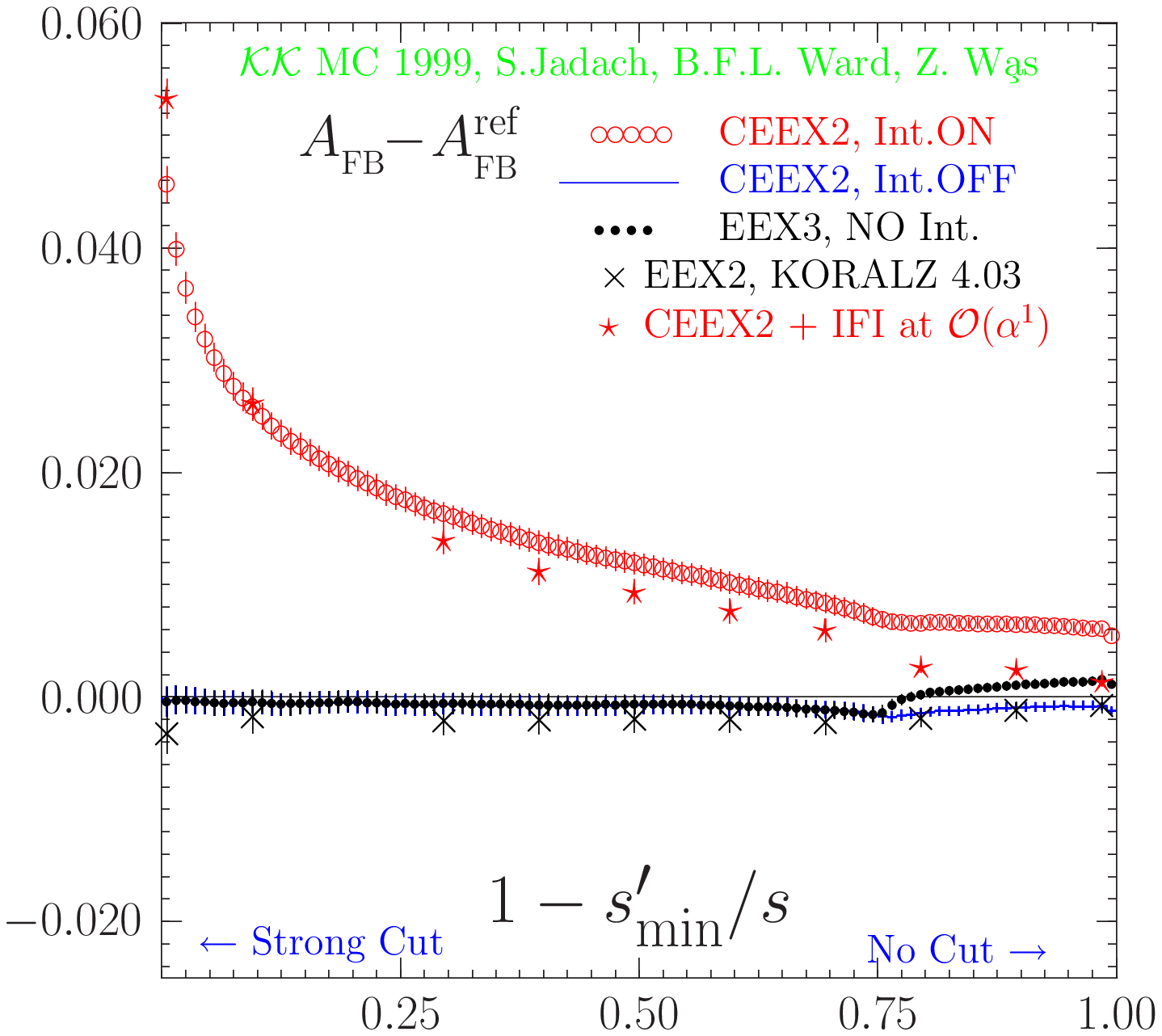,width=40mm,height=40mm}}}
\end{picture}
\caption{
{\Color{Orange}{\small\sf
  Absolute predictions for $\sigma_{tot},~A_{FB}$: $\mu\bar\mu,~189$ GeV.
}}}
\label{fig:2}
\end{center}
\end{figure}

\begin{figure}
\begin{center}
\setlength{\unitlength}{0.05mm}
\begin{picture}(800,1750)
\put(400,1750){\makebox(0,0)[b]{{\Color{Magenta} Physical Precision Tests}}}
\put( 400,1680){\makebox(0,0)[b]{ (a)}}
\put(400,800){\makebox(0,0)[b]{ (b)}}
\put(  -20,880){\makebox(0,0)[lb]{\epsfig{file=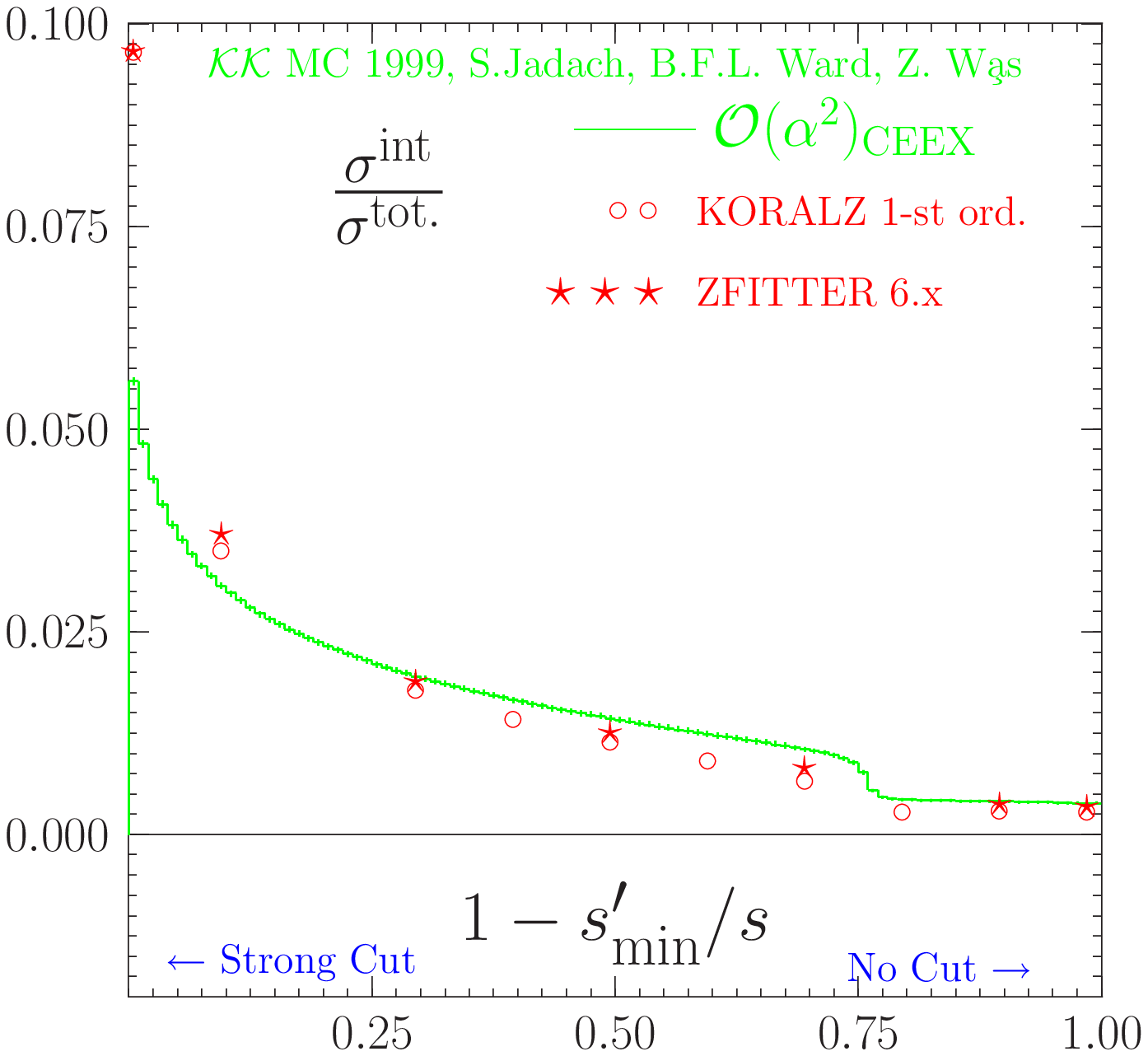,width=40mm,height=40mm}}}
\put(  -20, 0){\makebox(0,0)[lb]{\epsfig{file=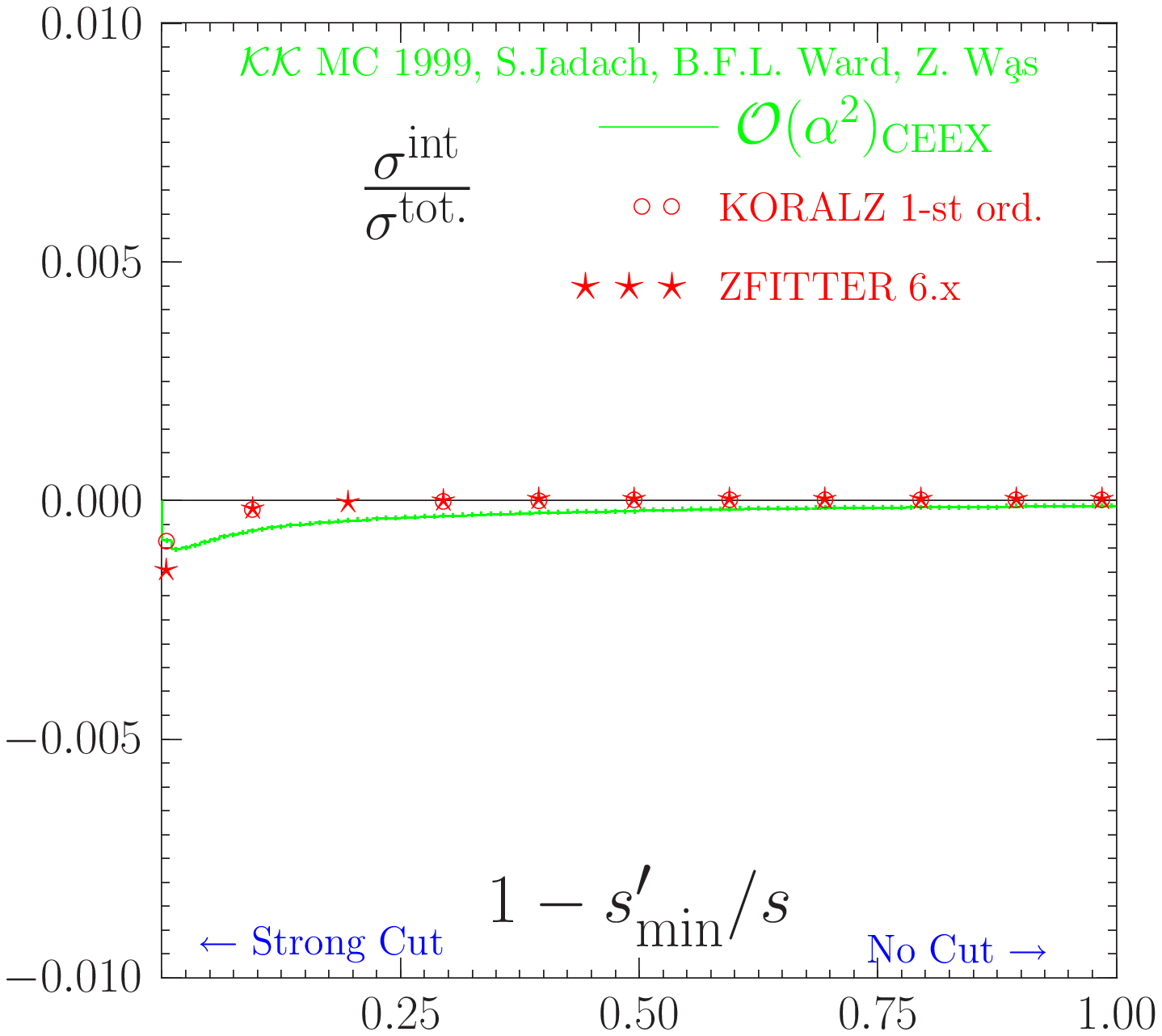,width=40mm,height=40mm}}}
\end{picture}
\caption{
{\Color{Orange}{\small\sf
  $s'$-cut dependence of $\delta\sigma$, No $\theta$-cut: (a), 189 GeV; (b), $M_Z$.
}}}
\label{fig:3}
\end{center}
\end{figure}

 In Figs.~\ref{fig:1},~\ref{fig:2} and \ref{fig:3} 
we show the baseline technical
precision test with the $\bar\beta_0$ level matrix element and
physical precision tests of $\sigma_{tot}$,~$A_{FB}$, and the IFI 
at LEP2 energies
as effected in the LEP2 MC Workshop~\cite{lp2mc-2}. 
With these and related tests we achieve
the technical precision tag of $0.02\%$ at LEP2 energies
, the physical tags of $0.2\%(0.2-0.4\%)$ for
the $\sigma_{tot}$(the $A_{FB}$), and firm control on the IFI~\cite{ceex}:
\begin{center} 
\begin{figure}
\epsfxsize140pt
\figurebox{140pt}{180pt}{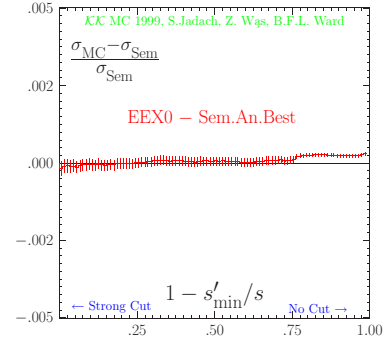}
\caption{Results for 189 GeV in the $\mu\bar\mu$ channel, 
for $v<0.999$. We plot the difference between the ${\cal KK}$ MC result
and semi-analytical (IEX) result divided by the latter.
}
\label{fig:1}
\end{figure}
\end{center}
we see that 
the IFI {\Color{Black}$\cong$} 1.5\% for {\Color{Red}energy cut} 0.3, that a 
{\Color{Black}$|cos\theta|<0.9$} cut reduces the IFI by {\Color{Red}25\%},
and that the IFI is 
{\Color{Red}very small} at the Z return, for example.
\par

\section{YFSWW3-1.14 MC}
Starting from the underlying process of interest, Eq.(\ref{eq1}),
its cross section, Eq.(\ref{eq2}),
{\bf\Color{Blue}
and the attendant 
{\Color{Red}$W^+W^-$ production and decay},
${\Color{Black}e^-(p_1) + e^+(p_2)} \to  {\Color{Red}W^-(q_1) + W^+(q_2)},~
{\Color{Red}W^-(q_1)} \to {\Color{Black}f_1(r_1)+\bar{f}_2(r_2)},~
 {\Color{Red}W^+(q_2)} \to {\Color{Black}f'_1(r'_1)+\bar{f}'_2(r'_2)}$,
we may isolate the Leading Pole Approximation 
{\Color{Maroon} (LPA$_{\Color{Magenta}a,b}$)} as follows:
{\small
\begin{equation}
  \label{eq:lpa}
  \begin{split}
   &{\cal M}^{(n)}_{{\Color{Red}4f}}({\Color{Black}p_1,p_2,r_1,r_2,r'_1,r'_2},{\Color{Orange}k_1,...,k_n})\;\;
    {{\Color{Maroon}LPA} \atop => }\\
&   {\cal M}^{(n)}_{{\Color{Red}LPA}}({\Color{Black}p_1,p_2,r_1,r_2,r'_1,r'_2},{\Color{Orange}k_1,...,k_n}) \\
   &=\sum_{{\Color{Orange}\gamma~Part'ns}}
   {\cal M}^{(n),{\Color{Red}\lambda_1 \lambda_2}}_{{\Color{Maroon}Prod}}({\Color{Black}p_1,p_2},{\Color{Red}q_1,q_2},{\Color{Orange}k_1,...,k_a})\\
   &\times {1\over {\Color{Red}D(q_1)}}\; 
       {\cal M}^{(n)}_{{\Color{Maroon}Dec_1},{\Color{Red}\lambda_1}}({\Color{Red}q_1},{\Color{Black}r_1,r_2},{\Color{Orange}k_{a+1},...,k_b})\; \\
    &\times {1\over {\Color{Red}D(q_2)}}\; 
       {\cal M}^{(n)}_{{\Color{Maroon}Dec_2},{\Color{Red}\lambda_2}}({\Color{Red}q_2},{\Color{Black}r'_1,r'_2},{\Color{Orange}k_{b+1},...,k_n}),
  \end{split}
\end{equation}
}
in an obvious notation~\cite{yfsww3} for the {\Color{Red}$W^{\pm}$} propagator
denominators ${\Color{Red}D(q_i)}$, etc.
Here, we can identify two different realizations,
LPA$_{{\Color{Magenta}a},{\Color{Black}b}}$,
of the leading
pole residues in Eq.~(\ref{eq:lpa}) by following the prescriptions of
{\Color{Green} Eden {\it et al.}~\cite{elop} and Stuart~\cite{stuart}:} in
${\cal M} = \sum_j {\Color{Black}\ell_j}{\Color{Magenta}A_j\left(\{q_kq_l\}\right)}$, the complete set 
of spinor covariants $\{\ell_j\}$ may (b) or may not (a)
be evaluated at the pole positions for the respective Lorentz scalar
functions $\{A_j\left(\{q_kq_l\}\right)\}$, as these latter
already realize the analyticity properties of the S-matrix by themselves. 
{\Color{Red} We do both.}

{\Color{Black}The standard YFS methods~\cite{eex}} {\Color{Red}(EEX-Type)}
give us the corresponding analog of Eq.(\ref{eq2}).
In realizing the exact ${\cal O}(\alpha)$ corrections in 
the latter equation 
in the LPA, 
we have chosen, for our renormalization scheme,
the {\Color{Magenta}$G_\mu$-Scheme} of Fleischer {\it et al.}~\cite{ew1}
in version 1.13 and the schemes {\Color{PineGreen}A}
and {\Color{Orange}B} in version 1.14, where in {\Color{PineGreen}A}
only the hard EW correction has {\Color{Magenta}$\alpha_{G_\mu}$}
whereas in {\Color{Orange}B} the entire {\Color{Magenta}${\cal O}(\alpha)$} 
correction has {\Color{Magenta}$\alpha(0)$}.
The analysis in Ref.~\cite{ward1} tells us that
the schemes A and B are improvements over the $G_\mu$ scheme in version
1.13, as we have verified in the context of the LEP2 MC Workshop
comparisons with Denner {\it et al.}
As a consequence, we have a
{\Color{Magenta}$-0.3\div-0.4\%$} shift of
the {\Color{Red}NORMALISATION} of version {\Color{Black}1.14}
relative to version {\Color{Black}1.13}.
{\Color{Green} See G. Passarino~\cite{passarino} for more
details and references.}
\par
\section{Results: YFSWW3-1.14}
In Fig.~\ref{fig:4}, we show the hardest photon angular distribution,
both at 200 GeV and at 500 GeV. We see that the NL EW correction
is relevant both for the BARE and CALO event selections as defined
Ref.~\cite{4f-report} away from the beam direction.
\begin{figure}
\begin{center}
\setlength{\unitlength}{0.05mm}
\begin{picture}(800,1600)
\put( 450,1520){\makebox(0,0)[cb]{$E_{CM} = 200\,GeV$} }
\put(450, 720){\makebox(0,0)[cb]{$E_{CM} = 500\,GeV$} }

\put(   0,800){\makebox(0,0)[lb]{
\epsfig{file=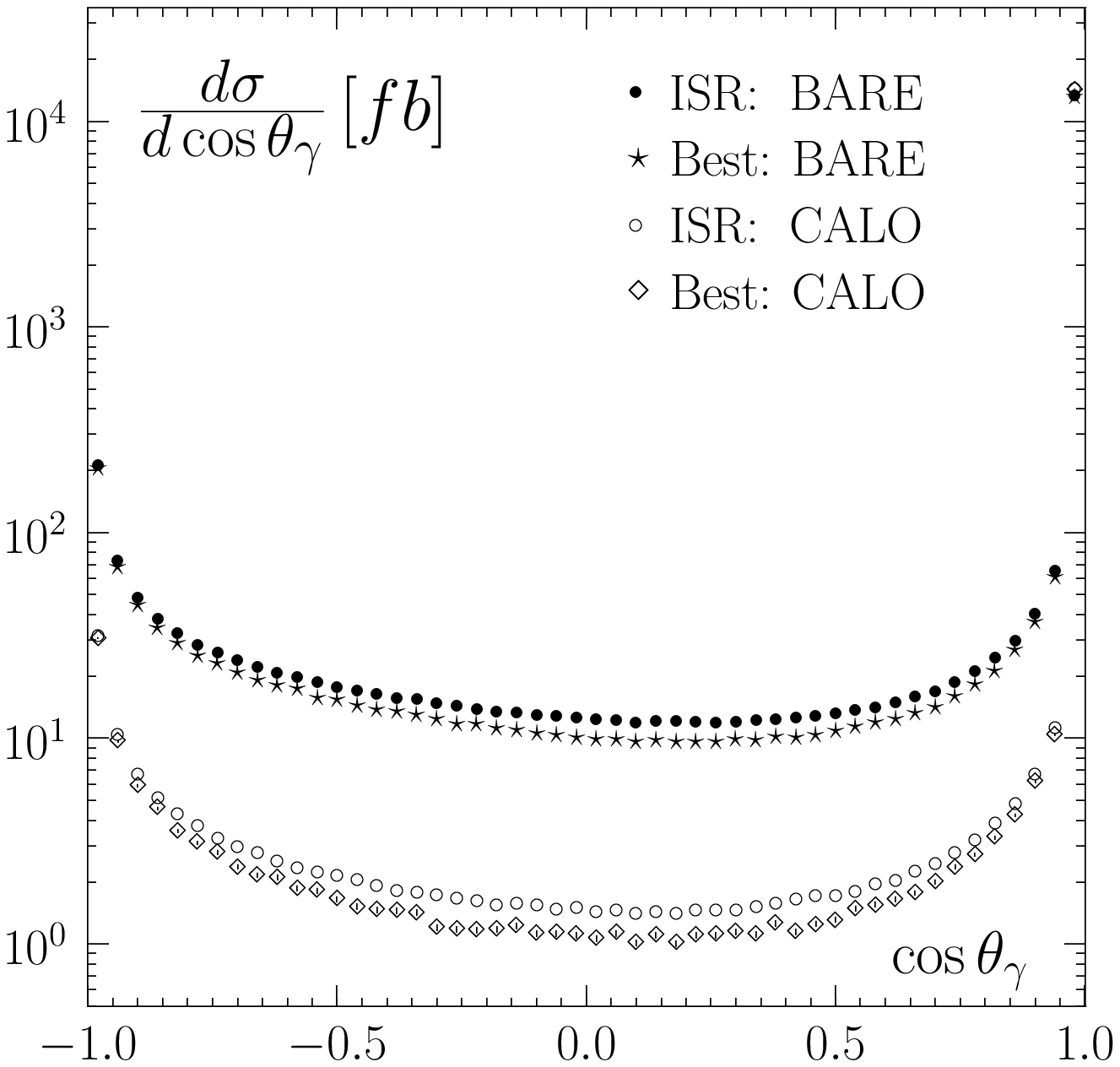       ,width=40mm,height=35mm}
}}

\put( 0,  0){\makebox(0,0)[lb]{
\epsfig{file=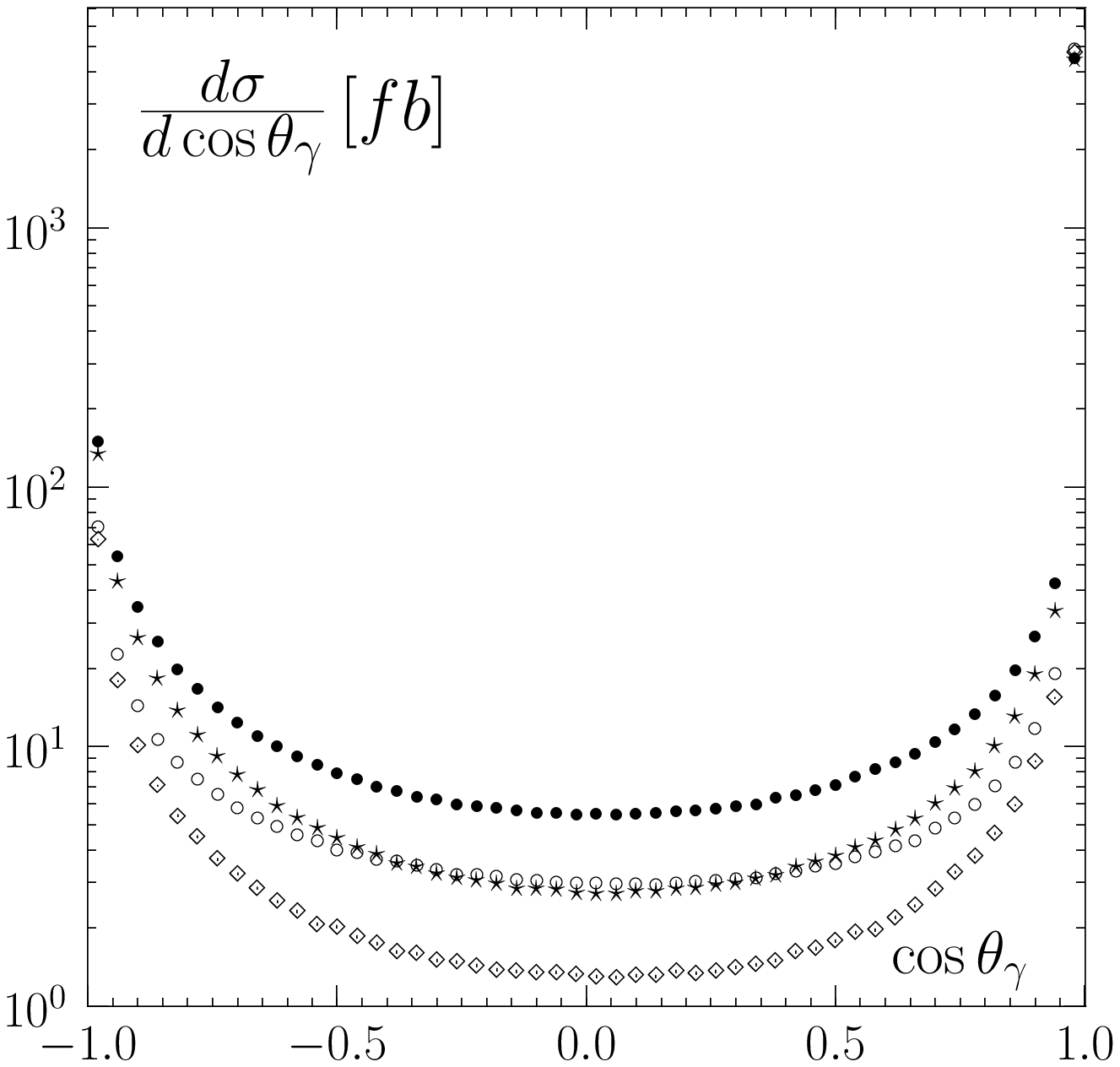       ,width=40mm,height=35mm}
}}

%
\end{picture}
\caption{
Distribution of {\Color{Orange}$cos\theta_\gamma$ {\Color{Black} 
with respect to the $e^+$ beam}}
}
\label{fig:4}
\end{center}
\end{figure}
Similar effects are discussed in Refs.~\cite{yfsww3}, where 
we find that the EW NL correction at LEP2 energies is large, $\sim -2\%$,
and is in general a non-trivial function of the kinematical variables.
The authors in Ref.~\cite{ditt} have reached the analogous conclusion.
\par
Indeed, in Table~\ref{tab:1} we show a comparison between the results
from RacoonWW and YFSWW3-1.14, where we have chosen the case
of without cuts, as carried out in the context of the LEP2 MC
Workshop. From these results and others similar to them we
arrive at the theoretical precision tag of $0.4\%$
at 200 GeV for the WW signal cross section at LEP2. See 
G. Passarino~\cite{passarino} for more details and references.
{\Color{Blue}  
\let\sstl=\scriptscriptstyle
\def\Was{W\c as}
\def\Order#1{${\cal O}(#1$)}
\def\Ordpr#1{${\cal O}(#1)_{prag}$}
\def\bbe{\bar{\beta}}
\def\tbe{\tilde{\beta}}
\def\tal{\tilde{\alpha}}
\def\tom{\tilde{\omega}}
\def\half{ {1\over 2} }
\def\alf1{ {\alpha\over\pi} }
\def\nl{\newline}
\def\Oaz{${\cal O}(\alpha^0)$}
\def\Oaf{${\cal O}(\alpha^1)$}
\def\Oas{${\cal O}(\alpha^2)$}
\begin{table}[t]
{\tiny
\begin{center}
\begin{tabular}{|c|c|c|c|}
\hline
\multicolumn{2}{|c|}{\bf {\Color{Red}no cuts}}&
\multicolumn{2}{|c|}{\bf{\Color{Black}$\sigma_{\mathrm{tot}}[\mathrm{fb}]$}}\\
\hline
final state & program & {\Color{Maroon}Born} & {\Color{PineGreen}best} \\
\hline\hline
& {\tt {\Color{Magenta}YFSWW3}} & {\Color{Magenta}219.770(23)} & {\Color{Magenta}199.995(62)} \\
$\nu_\mu\mu^+\tau^-\bar\nu_\tau$
& {\tt RacoonWW} & 219.836(40) & 199.551(46) \\
\cline{2-4}
& ({\Color{Magenta}Y}--R)/{\Color{Magenta}Y} & {\Color{Maroon}$-0.03(2)$\%} &  {\Color{PineGreen}0.22(4)\%} \\
\hline\hline
& {\tt {\Color{Magenta}YFSWW3}} & {\Color{Magenta}659.64(07)} & {\Color{Magenta}622.71(19)} \\
$\Pu\bar\Pd\mu^-\bar\nu_\mu$
& {\tt RacoonWW} & 659.51(12) & 621.06(14) \\
\cline{2-4}
& ({\Color{Magenta}Y}--R)/{\Color{Magenta}Y} & {\Color{Maroon}$0.02(2)$\%} &  {\Color{PineGreen}0.27(4)\%} \\
\hline\hline
& {\tt {\Color{Magenta}YFSWW3}} & {\Color{Magenta}1978.18(21)} & {\Color{Magenta}1937.40(61)} \\
$\Pu\bar\Pd\Ps\bar\Pc$
& {\tt RacoonWW} & 1978.53(36) & 1932.20(44) \\
\cline{2-4}
& ({\Color{Magenta}Y}--R)/{\Color{Magenta}Y} & {\Color{Maroon}$-0.02(2)$\%} &  {\Color{PineGreen}0.27(4)\%} \\
\hline
\end{tabular}
\end{center}
\caption{
{\Color{Orange}Total cross sections, CC03 from {\tt {\Color{Blue}RacoonWW}},
{\tt  {\Color{Magenta}YFSWW3}}, $\sqrt{s}=200\,\GeV$ 
{\Color{Red}without cuts}. Statistical errors -- last digits in $(~)$, etc.{\Color{PineGreen}$\Rightarrow 0.4\%$ TU.}}}
\label{tab:1}
}
\end{table}
}

\section{Conclusions}

Our conclusion for the CEEX ${\cal KK}$ MC discussion is that
the CEEX is a clear upgrade path for the EEX in a spin amplitude level MC.
We have shown that, for {\Color{Red}LEP2}, the total {\Color{Red}TU} is {\Color{Black}0.2\%}{\Color{Magenta}(0.2-0.4\%)} for {\Color{Black}$\sigma_{tot}$}{\Color{Magenta}$(A_{FB})$},
for typical cuts -- for the {\Color{Red}LC at 0.5 TeV}, 
these are a factor of 2 worse, and for
{\Color{Magenta}$\gamma\gamma^*$} the {\Color{Magenta}TU} is {\Color{Red}0.3\%} for {\Color{Black}LEP2} (there is no firm result for {\Color{Black}LC}).
The IFI {\Color{Magenta}(ISR$\otimes$FSR)} is included and under firm control. 
Our conclusions for YFSWW3-1.14 are that the {\Color{Magenta}EW NL} correction~\cite{ew1}
in ${\cal O}(\alpha)$, which is also realized in {\Color{Magenta}RacoonWW},
is important both for the normalisation and for the differential
distributions. The {\Color{Magenta}TU} at 200 GeV, based on comparisons with 
{\Color{Magenta}RacoonWW}, is {\Color{Magenta}0.4\%}.
\par

\section*{Acknowledgements}
The authors all thank the CERN TH Div. and the CERN
ALEPH, DELPHI, OPAL and L3 Collaborations for their
support and kind hospitality during the course of this work.
S. J. thanks the DESY TH. Div. for its support and kind hospitality
while a part of this work was done. This work was supported in part
by the Polish Government
grants KBN 2P03B08414 and 2P03B14715, the Maria Sk\l{}odowska-Curie
Joint Fund II PAA/DOE-97-316, and
by the US Department of Energy Contracts  DE-FG05-91ER40627
and   DE-AC03-76ER00515.

\end{document}